\documentclass[%
reprint,
amsmath,amssymb,
prb,
]{revtex4-1}

\usepackage{amsmath} 
\usepackage{amsthm} 
\usepackage{amssymb}	
 
\newcommand{\ket}[1]{\left| #1 \right>} 
\let\baraccent=\= 
\renewcommand{\=}[1]{\stackrel{#1}{=}} 

\theoremstyle{definition}

\theoremstyle{remark}

\newcommand{\degreeC}{^{\circ}\mathrm{C}} 

\usepackage{natbib}
\usepackage{graphicx}
\usepackage{dcolumn}
\usepackage{bm}

\begin{document}

\preprint{Preprint version: \today}

\title{Spin dependent recombination based magnetic resonance spectroscopy of bismuth donor spins in silicon at low magnetic fields}

\author{P. A. Mortemousque}
\affiliation{School of Fundamental Science and Technology, Keio University, 3-14-1 Hiyoshi, Kohoku-ku, Yokohama 223-8522, Japan}
\author{T. Sekiguchi}%
\affiliation{School of Fundamental Science and Technology, Keio University, 3-14-1 Hiyoshi, Kohoku-ku, Yokohama 223-8522, Japan}
\author{C. Culan}
\affiliation{School of Fundamental Science and Technology, Keio University, 3-14-1 Hiyoshi, Kohoku-ku, Yokohama 223-8522, Japan}


\author{M. P. Vlasenko}
\affiliation{A. F. Ioffe Physico-Technical Institute, Russian Academy of Sciences, 194021, St. Petersburg, Russia}%

\author{R. G. Elliman}
\affiliation{%
Australian National University, Research School of Physics and Engineering, Canberra, ACT 0200, Australia
}%

\author{L. S. Vlasenko}
\affiliation{A. F. Ioffe Physico-Technical Institute, Russian Academy of Sciences, 194021, St. Petersburg, Russia}%

\author{K. M. Itoh}
\affiliation{School of Fundamental Science and Technology, Keio University, 3-14-1 Hiyoshi, Kohoku-ku, Yokohama 223-8522, Japan}


\begin{abstract}
Low-field ($6-110$ mT) magnetic resonance of bismuth 
(Bi) donors in silicon has been observed by monitoring the change in photoconductivity 
induced by spin dependent recombination.
The spectra at various resonance frequencies show signal intensity distributions 
drastically different from that observed in conventional electron paramagnetic resonance, 
attributed to different recombination rates for the forty possible combinations of spin 
states of a pair of a Bi donor and a paramagnetic recombination center. 
An excellent tunability of Bi excitation energy for the future coupling with superconducting flux 
qubits at low fields has been demonstrated.
\end{abstract}

\maketitle


Among a variety of qubits investigated for the realization of solid-state
 quantum computers, superconducting qubits are the leading candidates for 
 quantum processors because of their fast operation capabilities ($\pi/2$ pulse 
 shorter than 10 ns).\cite{Chiorescu2003} However, the shortcoming of their 
 relatively fast decoherence time needs to be overcome by connecting to memory qubits 
 that can store quantum information throughout the course of computation. 
 This requires memory qubits working under low magnetic field, typically 
 below 10 mT for aluminum superconducting qubits,\cite{Cochran1958} since they 
 cannot operate at higher fields.

 In this context, the bismuth (Bi) donor in silicon (Si) has attracted much 
 attention recently. Its large hyperfine interaction $a/h=1.4754$ GHz 
 (Ref. [\onlinecite{Feher1959}]) 
 and the $^{209}$Bi nuclear spin $I=9/2$ give a large zero-field splitting 
 of 7.4 GHz. 
 This splitting is comparable to the typical energy splitting between 
 $\ket{R}$ and $\ket{L}$ states of superconducting flux qubits.\cite{Chiorescu2003,Zhu2011} 
 Thus, coupling between a Bi spin qubit and a superconducting flux qubit on Si 
 is in principle possible via a microwave photon through a waveguide.
 A proposal of such an application\cite{Morley2010} has prompted extensive research on the Bi 
 donor spins in Si very recently.\cite{Morley2010,George2010} 
 Starting from the spectroscopic analysis of the electron paramagnetic 
 resonance (EPR),\cite{George2010,Weis2012} the electron spin relaxation time $T_1$,\cite{Morley2010,Belli2011} 
 decoherence time $T_2$,\cite{Morley2010,Belli2011,Weis2012} and 
 superhyperfine interaction with nearby $^{29}$Si nuclear spins \cite{Belli2011} were 
 investigated. Moreover, the coherent transfer between electron and $^{209}$Bi nuclear 
 spins\cite{George2010} and dynamic nuclear polarization of $^{209}$Bi\cite{Morley2010,Sekiguchi2010} 
 were achieved.
 Yet all of these EPR studies were performed at 9 GHz (around 320 mT) 
 and at 240 GHz (around 8.6 T) excitation frequency. 

 In this paper we report on low-field ($6-110$ mT) radio frequency ($20-400$ MHz) and microwave 
 (8.141 GHz) magnetic resonance, as well as X-band (9 GHz) magnetic resonance, of ion-implanted 
 Bi donors in Si using a highly sensitive, spin dependent recombination based magnetic 
 resonance (SDR-MR) method.\cite{Vranch1988,Vlasenko1995,Laiho1995,Vlasenko1995a}
%
%

The samples were prepared from highly resistive ($>3000$ $\Omega\cdot$cm), float-zone 
(FZ) n-type silicon wafers implanted with Bi ions at room temperature with a 
total fluence of $2\times10^{12}$ cm$^{-2}$. 
The implantation energies are 300 and 550 keV with 
the doses of $0.7\times10^{12}$ and $1.3\times10^{12}$ cm$^{-2}$, 
respectively.  This condition leads to the Bi concentration of about $1.8\times10^{17}$ 
cm$^{-3}$ in the depth of 90 to 150 nm from the surface. The post-implantation 
annealing was performed at 650 $\degreeC$ for 30 min in an evacuated   
quartz tube. This annealing condition has been shown to achieve the 
electrical activation of 80 $\%$,\cite{Souza1993,Baron1969,Weis2012} 
thus yielding about $4.8\times10^{11}$ Bi donors in a 0.3 cm$^2$ sample area. 
%
%
%
The SDR-MR spectra were recorded at 16 K with a 
commercial continuous wave EPR spectrometer (JEOL JES-RE3X) working at X-band (9 GHz microwave)
with a homemade coil for radio frequency ($20-400$ MHz) and microwave (8.141 GHz) irradiation to 
induce magnetic resonance at low field ($6-110$ mT).  Continuous illumination with a 100-W 
halogen lamp generates photocarriers in the sample. Magnetic resonance 
can enhance the spin-dependent recombination, which decreases the 
density of photocarriers. Then, the absorption of the microwave 
electric field by the photocarriers is decreased, leading to an enhancement 
in the Q-factor of the cavity. Thus, the effect of magnetic resonance can be 
detected simply as the change in the X-band microwave reflection from the 
cavity. The second derivative of the reflected intensity with respect to the field modulation was recorded as an SDR signal to reduce the broad cyclotron resonance lines and the background change of the sample resistivity during the magnetic field scan. 
Note that because of our high power (80 mW) saturating excitation, i. e., making the populations of the ground- and excited-states the same, the conventional EPR absorption signal is suppressed. Such saturation is necessary to flip one of the spins in a pair of Bi and defect to induce SDR as we will discuss later.  

The spin system of an isolated Bi donor in static magnetic field $B_0$ (electron 
spin $S=1/2$ and $^{209}$Bi nuclear spin $I=9/2$) can be represented by the 
spin Hamiltonian:
\begin{equation}
\mathcal{H}=g\mu_BB_0S_z - g_n\mu_nB_0I_z+a\bm{S}\cdot\bm{I},
\end{equation}
\noindent where $\mu_B$ and $\mu_n$ are the Bohr and nuclear magnetons, and  
$g=2.0003$ (Ref. [\onlinecite{Feher1959}]) and $g_n=0.914$ (Ref. [\onlinecite{Morley2010}]) are Bi electron 
and nuclear $g$-factors, respectively. 
The SDR method requires 
a coupled pair of electron spins.\cite{Kaplan1978,Cox1978,Morishita2009} 
In the present study, the partner of the Bi donor electron spin is the 
electron spin of a deep paramagnetic recombination center (PRC) 
which is supposed to be created during the implantation process and not 
completely removed by controlling the annealing conditions.\cite{Miksic2002} We have attempted to identify the symmetry of the deep PRC by tracing the angular dependence of the EPR peaks.  However, the peaks were too broad to draw conclusions.  

%
\begin{figure}[b]
\includegraphics[width=7cm]{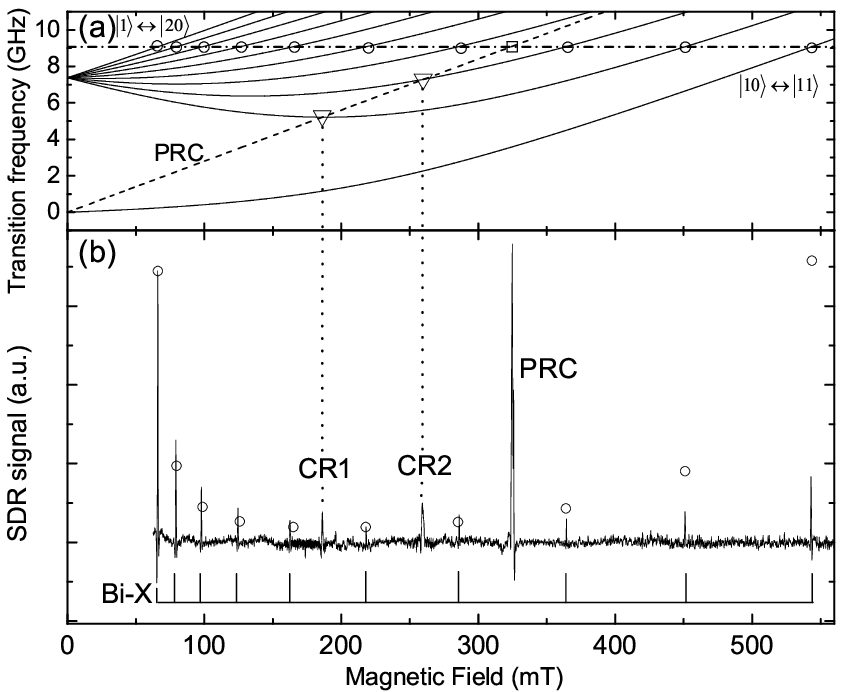}
\caption{\label{fig1} (a) Calculated EPR transition frequencies of Bi donors 
(solid curves) and deep paramagnetic recombination center (PRC) (dashed line) 
with 9.076 GHz microwave excitation frequency (dot-dashed line). The intersections of the PRC transition energy with bismuth donor transitions (open triangles) and with the 9.076 GHz microwave (open square) are also shown. 
(b) An SDR-MR spectrum of Bi donors in Si recorded at 16 K under illumination. The 9.076 GHz microwave is used both to induce Bi EPR transition and to probe the change in the sample photoconductivity. CR1 and CR2 are the cross relaxation signals between Bi and PRC. The open circles in (b) indicate simulated intensities using the SDR model described in the text with the parameter value of $R_\text{p}/R_\text{ap} = 0.01$. }
\end{figure}
The X-band (9.076 GHz) SDR-MR spectrum measured without radio frequency 
excitation is presented in Fig. 1(b). The peaks labeled as Bi-X and PRC  
indicate ten EPR-``allowed'' transitions of the Bi 
donors and one EPR transition of a PRC, respectively, corresponding to the intersections
with the 9.076 GHz excitation in Fig. 1(a).
Here the EPR transition frequencies of the Bi donor (solid curves)\cite{Mohammady2010} 
were calculated as functions of the static magnetic field $B_0$ using the 
spin Hamiltonian in Eq. (1) and that of PRC (dashed curve) was calculated as an 
isotropic, nuclear spin free, paramagnetic center $S=1/2$ and $g\approx2.005(3)$.
The same notation as in Ref. [\onlinecite{Mohammady2010}] for labeling Bi eigenstates is used; the labels 1 
to 20 in increasing order of energy.

In addition, two lines labeled as CR1 (186 mT) and CR2 (259 mT) arise 
due not to the resonance with the 9.076 GHz microwave but to cross 
relaxation (CR) between particular Bi donor transitions and the PRC 
transitions, in a way very similar to the cross relaxation between phosphorus 
donors and SL1 centers in Si observed by electrically detected magnetic 
resonance.\cite{Akhtar2011} This assignment of CR1 and CR2 is further 
justified in Fig. 2. Even with the decrease in the 
microwave excitation frequency, the position of these lines remains 
the same whereas the Bi EPR line positions shift to lower 
fields. This proves the presence of coupling between the Bi donor 
and PRC electron spins, which is requisite for the SDR detection 
method. 
%
\begin{figure}[b]
\includegraphics[width=7cm]{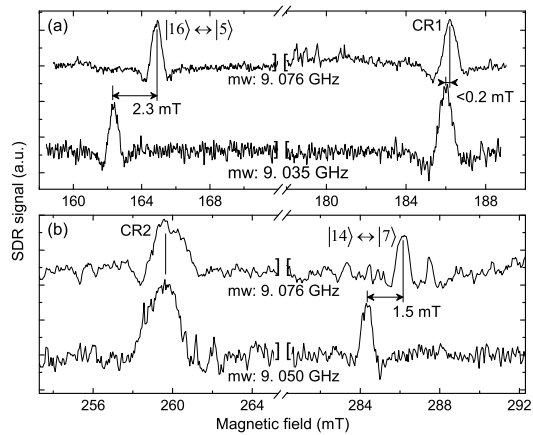}
\caption{\label{fig2} Cross relaxation lines CR1 (a) and CR2 (b) 
together with SDR-MR Bi lines detected at different microwave frequencies. 
The Bi EPR lines labeled as $\ket{16}\leftrightarrow\ket{5}$ in (a) 
and $\ket{14}\leftrightarrow\ket{7}$ in (b) shift with the resonant frequency 
whereas the lines CR1 and CR2 do not. Additionally, the CR1 line 
is narrower than CR2 because, as shown in Fig. 1(a), the difference in the field-derivative 
of the transition frequency between the paramagnetic recombination center and 
resonant Bi transition is larger.}
\end{figure}

We should emphasize that even at the same X-band resonance of the Bi donor, 
the observed SDR-MR line intensity distribution is clearly different from that 
observed in the conventional EPR spectra.
\cite{Morley2010,George2010,Mohammady2010,Belli2011} 
The intensity differs for the ten different transitions 
in the present SDR-MR whereas it is practically the same in conventional EPR, 
reflecting simply the thermal equilibrium population difference 
between the involved levels. Furthermore, the observed line-dependence of the SDR-MR 
intensity is distinctively stronger than the line dependence in the EPR transition.\cite{Mohammady2010}

Figure 3(a) shows the SDR-MR spectra probed by the same X-band microwave but with 
additional radio frequency excitation of 50 or 200 MHz. A simulation 
of the 200 MHz SDR-MR spectrum based on the SDR model that will be introduced 
later is also shown. Figure 3(b) shows the observed SDR-MR line positions 
(solid circles) for radio frequency excitation ranging from 20 to 400 MHz 
together with the calculated magnetic resonance transition frequencies for the Bi donor 
(solid curves). All of these simulated Bi magnetic resonance transitions are  
the $^{209}$Bi NMR transitions in the high field limit.
While the number of calculated Bi NMR lines in Fig. 3(b) appears  
ten, all but the lowest- and highest-field lines are nearly-degenerate 
doublets, i.e., each is composed of two lines separated by exactly 
twice the nuclear Zeeman splitting energy. Therefore, among the eighteen 
EPR transitions of the Bi donor excited by these radio frequencies, we 
observed clearly the two non-degenerate lines labeled as Bi-RF in Fig. 3. 
The remaining eight doublets are too weak to be observed with the current 
experimental conditions.
%
\begin{figure}[t]
\includegraphics[width=7cm]{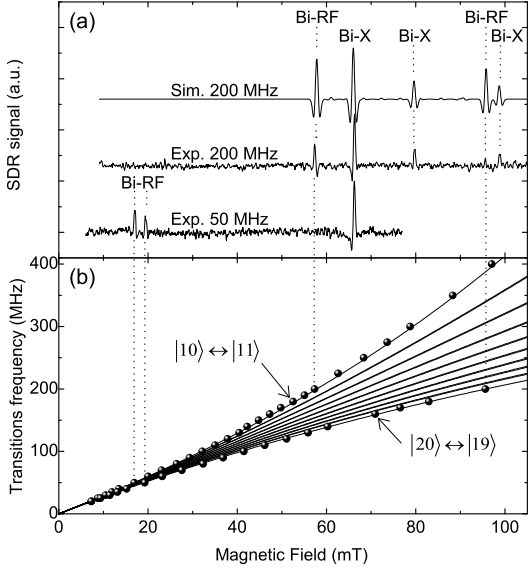}
\caption{\label{fig3} (a) Low-field SDR-MR spectra with 
50 and 200 MHz resonance frequencies together with simulation of the 
200 MHz spectrum. The transitions by the radio frequencies 
and X-band 9.076 GHZ microwave are labeled as Bi-RF and Bi-X, respectively. The 
line intensities are simulated using the same model and parameters as for Fig. 1(b).
(b) The Bi-RF line positions observed at various resonant frequencies (solid circles) 
together with calculated resonant fields (solid lines).}
\end{figure}

%
\begin{figure}[b]
\includegraphics[width=7cm]{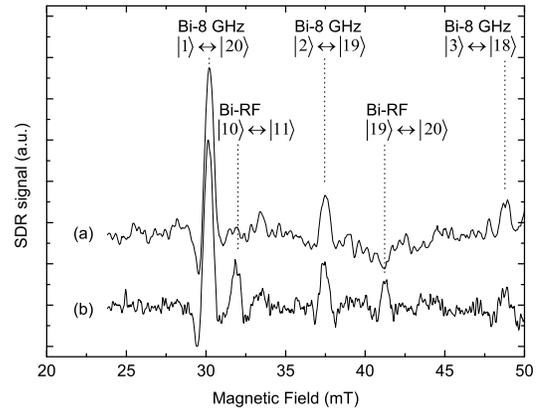}
\caption{\label{fig4} 
Low-field SDR-MR spectra probed by the 9.076 GHz X-band reflection (a) with a 
single 8.141 GHz excitation frequency and (b) with the same 8.141 GHz excitation 
plus an additional 100 MHz radio frequency. The lines resonant with the 
8.141 GHz microwave and the 100 MHz radio frequency are labeled as Bi-8 GHz and Bi-RF, 
respectively.}
\end{figure}

As mentioned above, considering only the EPR transition probabilities 
\cite{Mohammady2010} cannot describe the intensity of SDR-MR. 
In the following, we shall present a model, based on the SDR model developed in Refs. 
[\onlinecite{Boehme2001,Boehme2003,Boehme2006}], to simulate the SDR-MR spectra that are shown in Fig. 1(b) and Fig. 3(a). 
The SDR signal intensity, 
measured by probing the microwave intensity reflected by the cavity, is 
linear to the sample photoconductivity $\sigma$. 
Its change by magnetic resonance through SDR process can be written as:
\begin{equation}
\Delta\sigma\propto -\sum_{i,\mu}R_{i,\mu} \left[N_{i,\mu}(w\rightarrow\infty)-N_{i,\mu}(w=0)\right],
\end{equation}
\noindent where the subscripts $i$ and $\mu$ denote the Bi donor and PRC 
spin states, respectively. $R_{i,\mu}$ and $N_{i,\mu}$  are the recombination rate 
and population of the specified pair. Here the square bracket represents 
the change in the population from off-resonance ($w=0$) to saturated magnetic 
resonance ($w\rightarrow\infty$) conditions where $w$ is the excitation power. Eq. (2) 
is valid when the recombination rates can be assumed dominant over the pair generation 
and dissociation rates as well as the spin-lattice relaxation and spin decoherence rates. 
Furthermore, if only one transition between two Bi states 
$i$ and $j$ is selectively excited, the change in photoconductivity becomes:
\begin{eqnarray}
\Delta\sigma(i,j)\approx-\sum_{\mu} && [ R_{i,\mu} N_{i,\mu}(\infty) + R_{j,\mu} N_{j,\mu}(\infty)\nonumber\\
 && -(R_{i,\mu} N_{i,\mu}(0) + R_{j,\mu} N_{j,\mu}(0))].
\end{eqnarray}
\noindent Using the rate equations described in Ref. [17], Eq. (3) simplifies to:
\begin{equation}
\Delta\sigma(i,j)\approx -\sum_{\mu}\left[\frac{4}{R_{i,\mu}+R_{j,\mu}} - \left(\frac{1}{R_{i,\mu}}+\frac{1}{R_{j,\mu}}\right) \right].
\end{equation}
\noindent The first term of Eq. (4) corresponds to the number of recombining 
pairs when the resonance is saturated 
whereas the second and third terms are the off-resonance terms. 
Thus, a large change in SDR signal should be obtained when
either $R_{i,\mu}$ or $R_{j,\mu}$ is much smaller than the other. Finally, 
to evaluate the recombination rates $R_{i,\mu}$ and $R_{j,\mu}$, 
the product state of the Bi donor and the PRC is considered:
\begin{eqnarray}
\ket{i}\ket{\mu}=&&(\cos\phi_i(B_0)\ket{1/2,M-1/2}\nonumber\\
 &&+\sin\phi_i(B_0)\ket{-1/2,M+1/2})\ket{\mu}.
\end{eqnarray}
\noindent In the right-hand side, the Bi state ($i\in [1,20]$) is represented by the 
product of the electron ($m_S = \pm 1/2$) and nuclear spin 
($-9/2 \le m_I \le 9/2$) states with the total spin $z$-component of $-5\le M \le 5$ is represented 
on the basis of the electron ($m_S$) and nuclear ($m_I$) spin $z$-component eigenstates. 
\cite{Abragam1970} Note that there are two different Bi eigenstates for 
one particular $M$, except for $M=\pm5$. 
The mixing angle $\phi_i(B_0)$ depends on the Bi state and is a function of the parameters in the Hamiltonian, Eq. (1), as explicitly described in Ref. [\onlinecite{Mohammady2010}]. 
Depending on the state of PRC ($\mu=\pm1/2$, denoted by $\mu=\uparrow,\downarrow$), each term in Eq. (5) gives contribution 
to the recombination rate in terms of spin parallel ($R_\text{p}$) or anti-parallel 
($R_\text{ap}$) pair:
\begin{equation} 
R_{i,\uparrow}(B_0)=R_\text{p}\cos^2\phi_i(B_0) + R_\text{ap}\sin^2\phi_i(B_0), 
\end{equation}
\begin{equation}
R_{i,\downarrow}(B_0)=R_\text{p}\sin^2\phi_i(B_0) + R_\text{ap}\cos^2\phi_i(B_0).
\end{equation}
\noindent Then, only the recombination associated with the pure states, 
$i=10$ ($M=-5$) and $i=20$ ($M=+5$), have single components that are strictly 
parallel or anti-parallel, while the other states have a mixture of 
the two components. This, in combination with Eq. (4), is the reason why 
the highest- and lowest-field lines, which involve the Bi state $\ket{10}$ 
or $\ket{20}$, are stronger than the other lines at X-band resonance and 
exclusively strongest at the radio frequency resonance. 
We used this model to perform the simulation of the  
X-band spectrum in Fig. 1(b) and the 200 MHz spectrum in Fig. 3(a). The 
ratio $R_\text{p}/R_\text{ap}=0.01$ has led to good agreement with the 
experiments and is comparable to the recently reported value 
$15\,\mu\text{s}\,/\,2 \text{ ms} = 0.0075$ for the phosphorus donor in Si.\cite{Dreher2012} 
Note that, for the X-band spectrum we also take into account that an EPR transition line 
that is ``forbidden'' in the high-field limit overlaps with each EPR-``allowed'' one except for the 
highest- and lowest-field lines that involve one pure state.\cite{Mohammady2010}

Finally, we shall demonstrate the tunability of such ``pure-state'' transitions to the energy 
comparable to the superconducting qubits.  As shown in Fig. 4 (a), 
an additional 8.141 GHz microwave excitation in the same SDR method allows for 
successful excitation and detection of the EPR transition between Bi $\ket{1}$ and $\ket{20}$ 
levels at low field ($B_0 = 30$ mT).  Although it is preferred to achieve $B_0 <10$ mT for the 
coupling with the superconducting qubit, Fig. 4 shows clearly the flexibility to tune the 
energy difference between up and down states of the Bi electron spin. 
It is also possible to tune the superconducting flux qubit 
to match the energy between  $\ket{R}$ and $\ket{L}$ states 
with the Bi transition frequency separating $\ket{1}$ and $\ket{20}$ states. 
The coupling strength between the flux qubit 
and Bi is expected in the range of $1-100$ kHz.\cite{Zhu2011} 
Figure 4(b) shows a result of a similar experiment but with two excitation 
frequencies generated by two coils perpendicular to each other. The second coil 
was used to irradiate 100-MHz excitation frequency. The Bi-RF and Bi-8 GHz resonance 
lines are observed together. Hence, this experimental setup allows us to perform 
SDR-MR with two arbitrary excitation frequencies. 


In summary, we have obtained the electron paramagnetic resonance spectra of a small number  
($5\times10^{11}$) of Bi donors in Si 
at low magnetic field ($6-110$ mT). 
The detection was based on the measurement of the 
sample photoconductivity which changed significantly at the time of resonance due to 
specific spin-dependent-recombination phenomena.  
The spin-dependent-recombination process takes place via coupling of 
the electron spins between Bi donors and nearby paramagnetic recombination centers. 
The relative intensity of each resonance line has been described well by a 
spin-dependent-recombination model based on the mixing of Bi donor electron 
and nuclear spins.  

This work was supported in part by Grant-in-Aid for Scientific Research 
and Project for Developing Innovation Systems by MEXT, FIRST, and 
JST-EPSRC/SIC (EP/H025952/1).


\bibliography{20120811_manuscript_arXiv.bbl}

\end{document}